\documentclass[a4paper,aps,prd,twocolumn,nofootinbib]{revtex4}
\usepackage[latin9]{inputenc}
\usepackage{hyperref}
\hypersetup{
  colorlinks=true,        
  linkcolor=blue,         
  citecolor=cyan,         
}
\usepackage{breakurl}
\usepackage{graphicx}
\usepackage{epsf}
\usepackage{epsfig}
\usepackage{amssymb,amsmath}
\usepackage[usenames]{color}
\usepackage{amssymb}
\usepackage{times}
\usepackage{comment}

\newcommand\snn{\sqrt{s_\text{NN}}}

\newcommand\pt{p_\text{T}}

\newcommand\raa{R_\text{AA}}

\newcommand\dbar{\bar{D}^{0}}
\newcommand\dv {\Delta v_{1}}



\begin{document}

\title{Interactions between heavy quarks and tilted QGP fireballs in 200~AGeV Au+Au collisions}
\author{Ze-Fang Jiang$^{~1,2}$}
\email{jiangzf@mails.ccnu.edu.cn}
\author{Shanshan Cao$^{~3}$}
\email{shanshan.cao@sdu.edu.cn}
\author{Wen-Jing Xing$^{~3,2}$}
\author{Xiaowen Li$^{~3}$}
\author{Ben-Wei Zhang$^{~2,4}$}
\affiliation{$^1$ Department of Physics and Electronic-Information Engineering, Hubei Engineering University, Xiaogan, Hubei, 432000, China}
\affiliation{$^2$ Institute of Particle Physics and Key Laboratory of Quark and Lepton Physics (MOE), Central China Normal University, Wuhan, Hubei, 430079, China}
\affiliation{$^3$ Institute of Frontier and Interdisciplinary Science, Shandong University, Qingdao, Shandong, 266237, China}
\affiliation{$^4$ Guangdong Provincial Key Laboratory of Nuclear Science, Institute of Quantum Matter,
South China Normal University, Guangzhou, Guangdong, 510006, China}

\begin{abstract}
Heavy quark observables are applied to probe the initial energy density distribution with violation of longitudinal boost invariance produced in relativistic heavy-ion collisions. 
Using an improved Langevin model coupled to a (3+1)-dimensional viscous hydrodynamic model, 
we study the nuclear modification factor ($\raa$), directed flow ($v_1$) and elliptic flow ($v_2$) coefficients of heavy mesons and their decayed electrons at the RHIC energy.
We find that the counter-clockwise tilt of the nuclear matter in the reaction plane results in a positive (negative) heavy flavor $v_1$ in the backward (forward) rapidity region, whose magnitude increases with the heavy quark transverse momentum. The difference in the heavy flavor $\raa$ between different angular regions is also proposed as a complementary tool to characterize the asymmetry of the medium profile.
Our model results are consistent with currently available data at RHIC and provide predictions that can be tested by future measurements. 
\end{abstract}
\maketitle
\date{\today}

\section{Introduction}
\label{emsection1}

Heavy-ion collision experiments conducted at the BNL Relativistic Heavy-Ion Collider (RHIC) and the CERN Large Hadron Collider (LHC)  provide a unique opportunity to study the color deconfined state of nuclear matter, known as the Quark-Gluon Plasma (QGP)~\cite{Shuryak:2014zxa}. Heavy quarks serve as a clean probe that reveals the QGP properties at different energy scales~\cite{Dong:2019byy,Dong:2019unq}.
Due to the large mass of heavy quarks, they are mainly produced from the very early hard scatterings of high-energy nuclear collisions, prior to the formation of the QGP. Then they propagate through the medium and observe the entire evolution history of the QGP before they hadronize. Therefore, the difference in heavy flavor observables between proton-proton (p+p) and nucleus-nucleus (A+A) collisions characterizes the transport properties of the QGP~\cite{Cao:2018ews,Xu:2018gux}.

Considerable efforts have been devoted in developing transport models of heavy quarks in order to understand their dynamics inside a hot nuclear matter. It is now generally accepted that at high transverse momentum ($\pt$), perturbative calculations that involve both elastic and inelastic scatterings between heavy quarks and the QGP provide a successful description of the nuclear modification factor ($\raa$) of heavy flavor hadrons~\cite{Gossiaux:2010yx,Uphoff:2014hza,Nahrgang:2014vza,Cao:2015hia,Ke:2018tsh,Xing:2019xae,Li:2021xbd}.
At intermediate $\pt$, a combination of fragmentation and coalescence mechanisms is essential in understanding the hadronization process of heavy quarks thus describing the heavy flavor hadron chemistry observed at RHIC and LHC~\cite{Plumari:2017ntm,He:2019vgs,Cho:2019lxb,Cao:2019iqs}. At low $\pt$, modeling the non-perturbative scatterings between heavy quarks and the medium becomes inevitable in order to understand their strong interactions, as revealed by the large elliptic flow coefficients of $D$ mesons~\cite{He:2012df,Song:2015sfa,Das:2015ana,Scardina:2017ipo,Xing:2021xwc}. In addition to $\raa$ and $v_2$, novel observables have also been proposed to place more stringent constraints on the heavy quark dynamics inside the QGP, such as the momentum imbalance and angular correlation between heavy meson pairs~\cite{Nahrgang:2013saa,Cao:2015cba}, correlation of the higher-order harmonic flow coefficients between heavy and light flavor hadrons~\cite{Prado:2016szr}, and inner structures of heavy-flavor tagged jets~\cite{Dai:2022sjk,Wang:2019xey}.

The directed flow coefficient ($v_1$) of heavy quarks is another observable of great interest due to the copious information of the medium properties it encodes. It was proposed that due to the asymmetric distribution of the nuclear matter along the longitudinal direction, the heavy meson $v_1$ could be more than an order of magnitude larger than that of the light flavor hadrons emitted from the QGP~\cite{Chatterjee:2017ahy,Chatterjee:2018lsx,Nasim:2018hyw}. This has soon been confirmed by the STAR measurement~\cite{STAR:2019clv} and attracted many further studies~\cite{Oliva:2020doe,Beraudo:2021ont,Jiang:2022uoe} that couple various transport models to a tilted QGP fireball in the reaction plane~\cite{Bozek:2010bi}. Meanwhile, the splitting of $v_1$ ($\dv$) between heavy quarks and their anti-particles is also considered an effective tool to probe the extremely strong electromagnetic field generated by non-central heavy-ion collisions because of the opposite Lorentz force exerted on them~\cite{Das:2016cwd,Chatterjee:2017ahy,Chatterjee:2018lsx,Oliva:2020doe,Sun:2021joa,Jiang:2022uoe}. Interestingly, while a decreasing $v_1$ with respect to rapidity ($y$) is observed for both $D^{0}$ and $\dbar$ at STAR~\cite{STAR:2019clv}, with small difference between them, apparent splitting of $v_1$ is seen by ALICE~\cite{ALICE:2019sgg}, with $D^{0}$ increasing but $\dbar$ decreasing with respect to pseudorapidity ($\eta$). This puzzling observation implies the competing effects between the longitudinally tilted medium geometry and the electromagnetic field on the heavy flavor $v_1$ at RHIC and LHC.

In our previous study~\cite{Jiang:2022uoe}, it has been found that while the formation of heavy flavor $v_1$ is dominated by the deformed medium profile at the RHIC energy, it is mainly determined by the electromagnetic field at LHC. As a follow-up study, we will focus on the 200~AGeV Au+Au collisions at RHIC in the present work and conduct a systematic exploration on how heavy quarks can be utilized to probe the initial energy density distribution of the QGP. In addition to $D$ mesons, $\raa$, $v_1$ and $v_2$ will also be calculated for $B$ mesons and their decayed electrons. We will study the transverse momentum dependence of the heavy flavor $v_1$, and extract the slope parameter of the $v_1(y)$ function, which can be tested by future more precise measurements and help quantify the tilt of the QGP in its initial state. Last but not least, the difference in the heavy flavor $\raa$ between different angular regions will also be investigated as an alternative tool to characterize the asymmetry of the medium along different directions.

This paper will be organized as follows. In Sec.~\ref{section2}, we will provide a brief overview of our model setup, including a tilted initial condition of the bulk medium with respect to the longitudinal direction and its evolution via the CLVisc hydrodynamic model in Sec.~\ref{subsec:QGP}, and a modified Langevin approach that describes the heavy quark interaction with the QGP in Sec.~\ref{subsec:Langevin}. In Sec.~\ref{section4} we will present our numerical results on the heavy flavor $\raa$, $v_{1}$ and $v_{2}$, and study how they depend on the heavy quark mass, transverse momentum and medium geometry. In the end, we summarize and discuss future developments in Sec.~\ref{section5}.

\section{Heavy quark interaction with the QGP}
\label{section2}

\subsection{Hydrodynamic simulation with tilted initial condition}
\label{subsec:QGP}

In this work, the spacetime evolution profile of the QGP is calculated using the (3+1)-dimensional viscous hydrodynamic model CLVisc~\cite{Pang:2012he,Pang:2018zzo,Wu:2018cpc,Wu:2021fjf}. The initial energy density distribution is modeled with a parameterization that takes into account a tilt of the medium produced by non-central heavy-ion collisions~\cite{Jiang:2021foj,Jiang:2021ajc}. Its dependence on the transverse coordinates $(x,y)$ and the spacetime rapdity $(\eta_\mathrm{s})$ is given by
\begin{equation}
\begin{aligned}
\varepsilon(x,&y,\eta_\mathrm{s})= K \cdot \frac{0.95W_\text{N}(x,y,\eta_\mathrm{s})+0.05 n_\text{BC}(x,y)}{\left[0.95W_\text{N}(0,0,0)+0.05 n_\text{BC}(0,0)\right]|_{\mathbf{b}=0}}\\
                         &\times \exp\left[-\frac{(|\eta_\mathrm{s}|-\eta_\mathrm{w})^{2}}{2\sigma^{2}_{\eta}}\theta(|\eta_\mathrm{s}|-\eta_\mathrm{w}) \right],
\label{eq:ekw}
\end{aligned}
\end{equation}
where $K$ is an overall normalization factor that is fixed by the multiplicity distribution of the final charged particles ($dN_{\textrm{ch}}/d\eta$) observed in experiments, $n_\text{BC}$ is the distribution of binary collision points from the Glauber model, $\mathbf{b}$ represents the impact parameter, and $W_\text{N}$ is the distribution of wounded nucleons parameterized as
\begin{equation}
\begin{aligned}
W_\text{N}(x,y,\eta_\mathrm{s})=&[T_{1}(x,y)+T_{2}(x,y)]\\
+&H_\mathrm{t}[T_{1}(x,y)-T_{2}(x,y)]\tan\left(\frac{\eta_\mathrm{s}}{\eta_\mathrm{t}}\right).
\label{eq:mnccnu}
\end{aligned}
\end{equation}
Here, $T_1(x,y)$ and $T_2(x,y)$ are the density distributions of participant nucleons from the projectile and target nuclei propagating along the positive and negative longitudinal ($z$) directions respectively, and $H_\mathrm{t}\tan (\eta_\mathrm{s}/\eta_\mathrm{t})$ is introduced to model the imbalance of hadron emission between forward and backward rapidities.
In addition, at the end of Eq.~(\ref{eq:ekw}), an envelope function in the Gaussian form is used to
describe the plateau structure of the hadron yield observed at mid-rapidity, in which $\eta_\mathrm{w}$ is the width of the central rapidity plateau and $\sigma_{\eta}$ controls the speed of decay away from the plateau region~\cite{Pang:2018zzo}.
In Tab.~\ref{t:modelparameters}, we summarize all related model parameters introduced above.
These values have been adjusted in Ref.~\cite{Jiang:2021ajc} for a satisfactory description of the light hadron yield $dN_{\mathrm{ch}}/d\eta$ and their directed flow coefficient $v_{1}$ measured at RHIC and LHC.

\begin{figure}[tbp]
\begin{center}
\includegraphics[trim=0cm 0.0cm 0cm 0cm,width=8.0 cm,height=6 cm,clip]{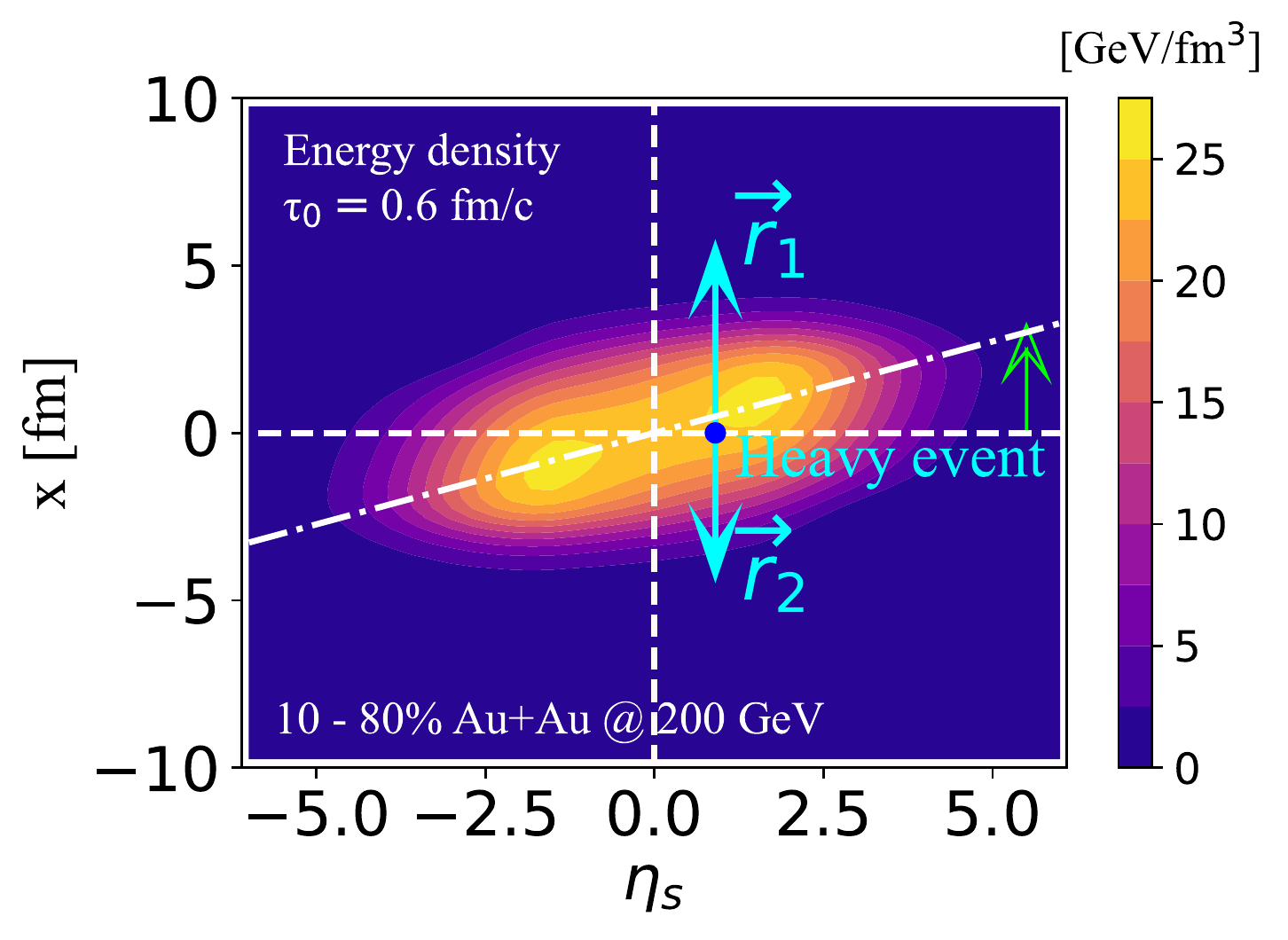}~~~\\
\includegraphics[trim=0cm 0.0cm 0cm 0cm,width=8.0 cm,height=6 cm,clip]{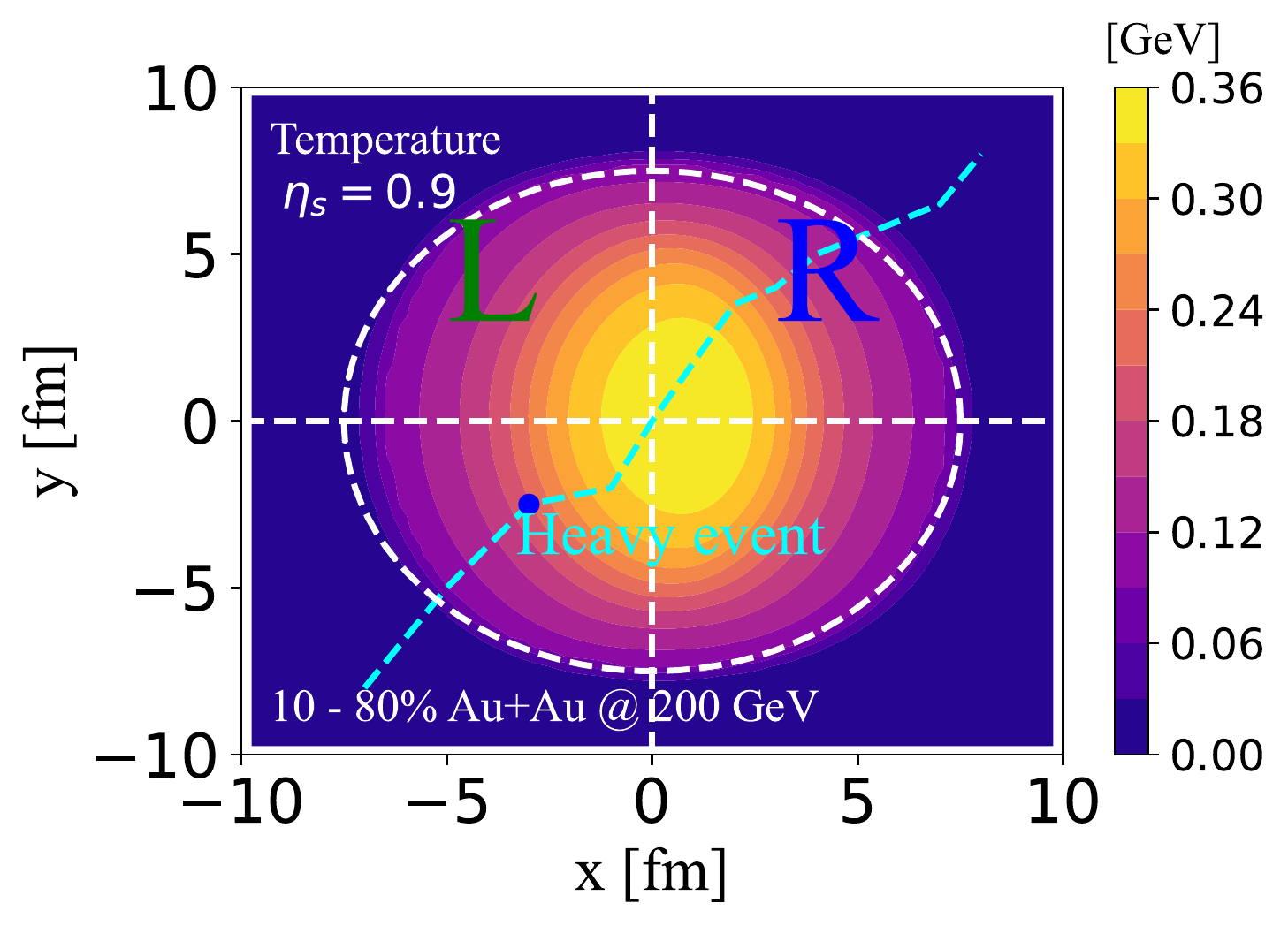}
\end{center}
\caption{(Color online) The initial profile of the QGP at $\tau_0 = 0.6$~fm/$c$ for 10-80\% ($b = 8.5$~fm) Au+Au collisions at $\snn=200$~GeV, upper panel for the side view of the energy density distribution, and lower panel for the top view of the temperature distribution. The solid arrows $\vec{r}_{1}$ and $\vec{r}_{2}$ (aqua color) sketch the heavy quark propagation, and the empty arrow (limes color) denotes the counter-clockwise tilt of the medium in the $\eta_\mathrm{s}$-$x$ plane.}
\label{fig:tiltedMedium}
\end{figure}

\begin{table}[!h]
\begin{center}
\begin{tabular}{| c| c |c| c| c| c | c |}
\hline
\hline
$\tau_{0}$ (fm/$c$)             & $K$ (GeV/fm$^{3}$)        &~~$\eta_\mathrm{w}$ ~~ & ~~$\sigma_{\eta}$ ~~  &~~$H_{\textrm{t}}$ ~~ &~~$\eta_\mathrm{t}$ ~~ &$ T_{\textrm{frz}}~(\textrm{MeV})$       \\
\hline
 0.6                   & 35.5       & 1.3     & 1.5  &3.9  &8.0  &137 \\
\hline
\hline
\end{tabular}
\caption{\label{t:modelparameters} Model parameters of the initial condition and hydrodynamic evolution for 10-80\% ($b = 8.5$~fm) Au+Au collisions at $\snn$ = 200 GeV~\cite{Pang:2018zzo,Loizides:2017ack}.}
\end{center}
\end{table}

The initial fluid velocity is assumed to follow the Bjorken approximation, where $v_{x} = v_{y} =0$ and $v_{z} = z/t$~\cite{Jiang:2021foj}. The initial transverse expansion and the asymmetric distribution of $v_z$ along the impact parameter ($x$) direction are neglected in the present study, though the latter will become crucial when discussing the development of global polarization in heavy-ion collisions~\cite{Li:2022pyw}.

With these setups, we present the 3-dimensional profile of the nuclear matter at the initial time of hydrodynamic evolution $\tau_0=0.6$~fm/$c$ in Fig.~\ref{fig:tiltedMedium}, the upper panel for the side view ($\eta_\mathrm{s}$-$x$ plane) of the energy density distribution, and the lower panel for the top view ($x$-$y$ plane) of the temperature distribution. In the figure, a clear counter-clockwise tilt of the medium in the $\eta_\mathrm{s}$-$x$ with respect to the longitudinal direction ($\eta_\mathrm{s}$) can be seen, which was shown to be essential for understanding the non-zero directed flow of soft hadrons emitted from the QGP~\cite{Jiang:2021ajc}. 
On the other hand, since heavy quarks are produced in the initial hard scatterings of nuclear collisions, they are expected to distribute symmetrically around the center $(0,0)$ of the overlapping region between the two colliding nuclei. As a result, they propagate through different path length, thus suffer different amount of energy loss, towards different directions at finite rapidity. For instance, at forward rapidity ($y>0$), heavy quarks traverse longer path length towards $+x$ (right) than $-x$ (left), resulting in a negative $x$-component of the average heavy quark momentum ($\langle p_x\rangle$) in the end. 

With the tilted initial condition above, we use the CLVisc hydrodynamic model to evolve the QGP profile. The hydrodynamic equation reads
\begin{equation}
\partial_{\mu}T^{\mu\nu}=0,
\label{eq:tmn}
\end{equation}
where the energy-momentum tensor $T^{\mu\nu}$ is given by
\begin{equation}
T^{\mu\nu}=\varepsilon u^{\mu}u^{\nu}-(P+\Pi)\Delta^{\mu\nu} + \pi^{\mu\nu},
\label{eq:tensor}
\end{equation}
with $\varepsilon$ being the local energy density,
$u^{\mu}$ being the fluid four velocity, $P$ being the pressure, $\Pi$ being the bulk viscosity pressure and $\pi^{\mu\nu}$ being the shear viscosity tensor. In addition,
$\Delta^{\mu\nu} = g^{\mu\nu}-u^{\mu}u^{\nu}$ is the projection operator with the metric tensor $g^{\mu\nu} = \text{diag} (1,-1,-1,-1)$.
In this study, the shear-viscosity-to-entropy-density-ratio is set as $\eta_\mathrm{v}/s = 0.08$ ($\eta_\mathrm{v}$ for the shear viscosity), while the bulk viscosity and the net baryon density are ignored. The equation of state (EoS)
is taken from the Wuppertal-Budapest work~\cite{Borsanyi:2013bia}. After hydrodynamic evolution, the QGP medium is converted to light flavor hadrons according to the Cooper-Frye mechanism with the isothermal freeze-out condition determined by a constant temperature $T_{\textrm{frz}}=137$~MeV.
These setups allow a reasonable description of the soft hadron spectra and their directed and elliptic flow coefficients observed at RHIC and LHC~\cite{Pang:2018zzo,Jiang:2021ajc,Jiang:2021foj}.

\subsection{Transport of heavy quarks}
\label{subsec:Langevin}

The interactions between heavy quarks and the QGP medium is described using our modified Langevin approach~\cite{Cao:2013ita,Cao:2015hia} that includes both elastic and inelastic scattering processes. 
The modified Langevin equation reads
\begin{equation}
\begin{aligned}
\frac{d\vec{p}}{dt}=-\eta_\mathrm{D}(\vec{p})\vec{p}+\vec{\xi}+\vec{f}_{g}, 
\label{eq:CaoLangevin}
\end{aligned}
\end{equation}
where $-\eta_\mathrm{D}(\vec{p})\vec{p}$ provides the drag force and $\vec{\xi}$ gives the thermal random force on heavy quarks inside a thermal medium. The third term $\vec{f}_{g}$ is introduced to describe the recoil force experienced by heavy quarks when they emit medium-induced gluons.

For quasielastic scatterings, we assume that $\vec{\xi}$ is independent of momentum ($\vec{p}$) in the present work. Its strength is determined by the white noise $\big\langle \xi^i (t) \xi^j(t') \big\rangle = \kappa \delta^{ij} \delta(t-t')$ where $\kappa$ is the momentum space diffusion coefficient of heavy quarks. Here $\kappa$ is further related to the drag coefficient via the fluctuation-dissipation relation $\eta_\mathrm{D}(p)=\kappa/(2TE)$, with $T$ and $E$ being the medium temperature and the heavy quark energy respectively. 
The spatial diffusion coefficient of heavy quarks can then be extracted as $D_\mathrm{s}\equiv T/[M\eta_\mathrm{D}(0)]=2T^2/\kappa$ in which $M$ is the heavy quark mass. This $D_\mathrm{s}$ will be treated as the only model parameter for our modified Langevin approach~\cite{Cao:2013ita,Cao:2015cba}. For a minimal model, a constant value of $D_\mathrm{s}(2\pi T)$ is used in this work, which is determined by the $\raa$ of heavy mesons and their decayed electrons, as will be shown in the next section. A more elaborate dynamical calculation of this diffusion coefficient has been developed in our recent work based on a non-perturbative potential scattering approach~\cite{Xing:2021xwc}, which can also be implemented in this Langevin model in our future study.

The recoil force in Eq.~(\ref{eq:CaoLangevin}) is given by $\vec{f_g}=-d\vec{p}_g/dt$, where $\vec{p}_g$ denotes the momentum of medium-induced gluons, whose spectrum can be taken from the higher-twist energy loss calculation~\cite{Guo:2000nz,Majumder:2009ge,Zhang:2003wk}. 
The strength of this term is characterized by the jet quenching parameter $\hat{q}$, 
which can be directly related to the momentum space diffusion coefficient of heavy quarks via a dimension factor -- $\hat{q}=2\kappa$~\cite{Cao:2015hia} -- and is further connected to the $D_\mathrm{s}$ parameter as discussed earlier.

For heavy quark production and evolution in realistic heavy-ion collisions, we initialize the spatial distributions of heavy quarks using the Monte-Carlo Glauber model, and their momentum spectra using the Fixed-Order-Next-to-Leading-Log (FONLL) perturbative QCD calculation~\cite{Cacciari:2001td,Cacciari:2012ny,Cacciari:2015fta} that includes both pair production and flavor excitation processes. In this study, the FONLL calculation is coupled to the CT14NLO parton distribution function (PDF)~\cite{Kretzer:2003it} and the EPPS16 parametrization~\cite{Eskola:2009uj} of nuclei to take into account the nuclear shadowing effect in heavy-ion collisions.
We assume the interactions between heavy quarks and the QGP start from the initial time ($\tau_0=0.6$~fm/$c$) of the hydrodynamic evolution of the nuclear matter. During the QGP stage, the energy-momentum of heavy quarks are updated according to Eq.~(\ref{eq:CaoLangevin}) in the local rest frame of the QGP. The local temperature and flow velocity information of the QGP are provided by the CLVisc hydrodynamic model as described in the previous subsection. 
When heavy quarks travel across the QGP boundary, defined by a hypersurface at a fixed decoupling temperature $T_\text{d} = 165$~MeV in this work, they are converted to heavy flavor hadrons via a hybrid fragmentation and coalescence model~\cite{Cao:2019iqs} that is well constrained by the heavy flavor hadron chemistry measured at RHIC and LHC. Finally, the heavy flavor hadrons decay into electrons via Pythia simulation~\cite{Sjostrand:2006za}. For a summary on the systematic uncertainties contributed by various ingredients of our model, one may refer to Ref.~\cite{Li:2020kax}.

\section{The nuclear modification factor and collective flow coefficients}
\label{section4}
In this section, we provide calculations on the nuclear modification factors and collective flow coefficients of heavy flavor mesons and their decayed electrons, and discuss how they are affected by a tilted QGP fireball. The nuclear modification factor ($R_\mathrm{AA}$) is defined as the ratio of particle spectra between A+A and p+p collisions, normalized with the average number of binary collisions ($\mathcal{N}_\mathrm{coll}$) per A+A collision:
\begin{equation}
\begin{aligned}
R_{\textrm{AA}}(y,p_{\textrm{T}},\phi_p)=\frac{1}{\mathcal{N}_\mathrm{coll}}\frac{dN_{\textrm{AA}}/dydp_{\textrm{T}}d\phi_p}{dN_{\textrm{pp}}/dydp_{\textrm{T}}d\phi_p}.
\label{eq:raa}
\end{aligned}
\end{equation}
For collective flow coefficients, we focus on the directed flow 
\begin{equation}
\begin{aligned}
v_{1}=\left\langle \cos(\phi-\Psi_1)\right\rangle=\left\langle\frac{p_{x}}{\pt}\right\rangle,
\label{eq:v1}
\end{aligned}
\end{equation}
and the elliptic flow
\begin{equation}
\begin{aligned}
v_{2}=\left\langle \cos(2(\phi-\Psi_2))\right\rangle=\left\langle\frac{p_{x}^{2}-p_{y}^{2}}{p_{x}^{2}+p_{y}^{2}}\right\rangle,
\label{eq:v1}
\end{aligned}
\end{equation}
in the present study, which can be viewed as the first and second order Fourier coefficients of the angular distribution of the particle spectra respectively. In the above equations, $\Psi_1$ and $\Psi_2$ represent the first-order and the second-order event plane angles and $\langle...\rangle$ denotes average over both the final-state particles and different collision events. Since we use the modified optical Glauber model to calculate the initial energy density distribution of the QGP (as described in Sec.~\ref{section2}), the event-by-event fluctuations have not been taken into account in this work~\cite{Cao:2015cba,Jiang:2022uoe}. Therefore, the event plane in the final state is the same as the participant plane in the initial state and also the spectator plane that can be measured from the deflected neutrons in experiments~\cite{Jiang:2022uoe}. 

\subsection{$\raa$, $v_{1}$ and $v_{2}$ of heavy mesons}
\label{section4-1}

\begin{figure}[tbp]
\begin{center}
\includegraphics[trim=0cm 0.2cm 0cm 0cm,width=8 cm,height=5 cm,clip]{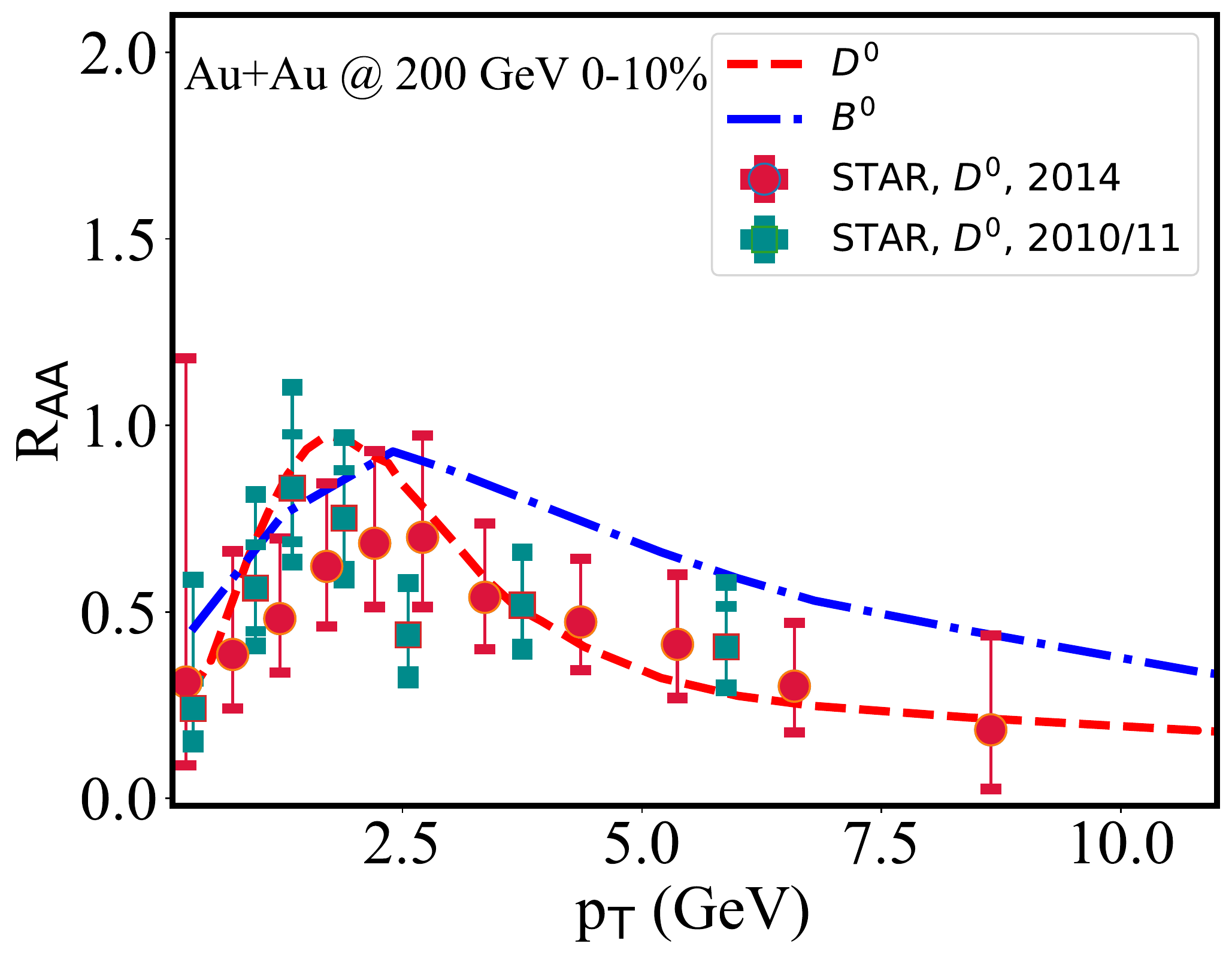}~\\
\includegraphics[trim=0cm 0.2cm 0cm 0cm,width=8 cm,height=5 cm,clip]{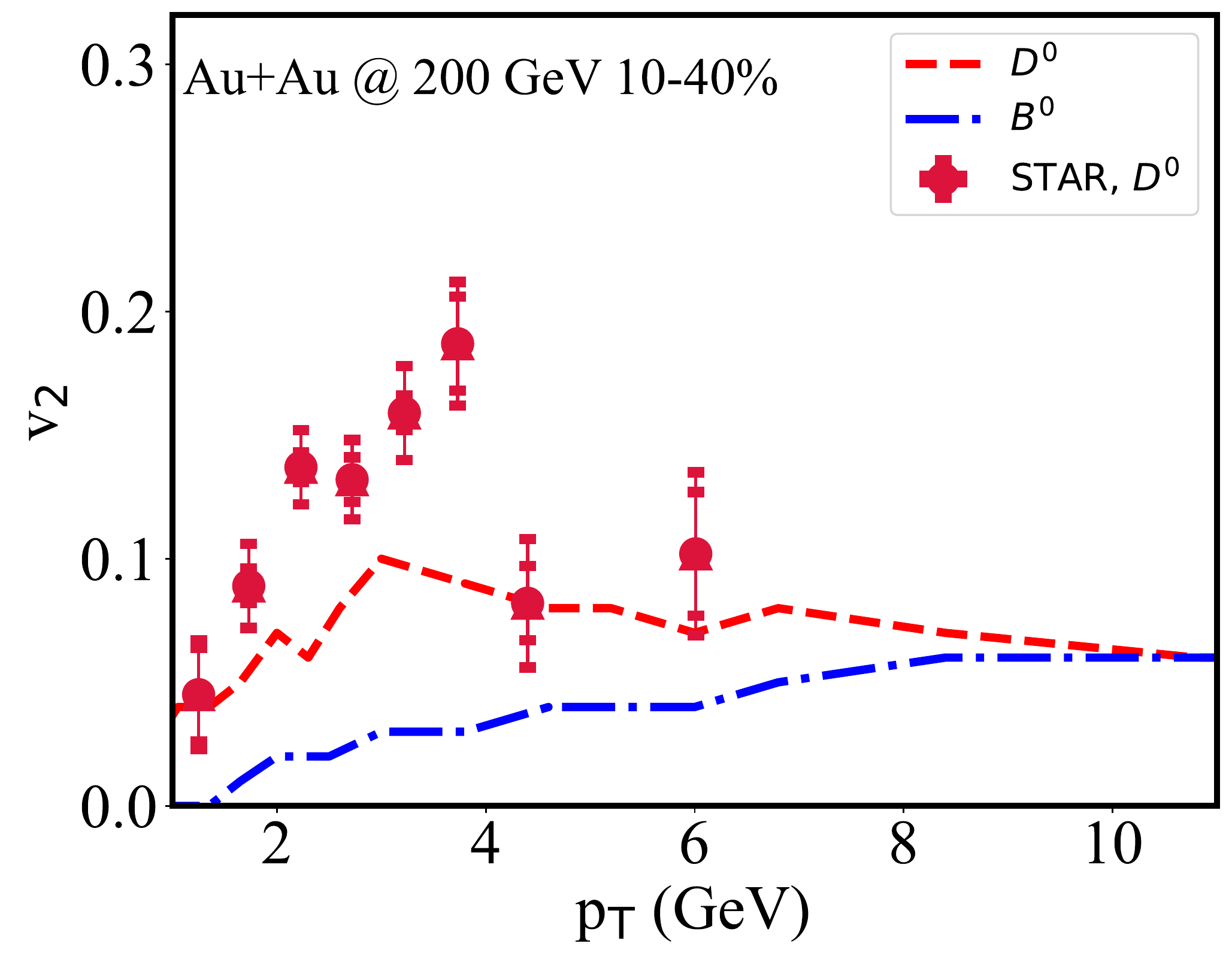}~\\
\includegraphics[trim=0cm 0.2cm 0cm 0cm,width=8 cm,height=5 cm,clip]{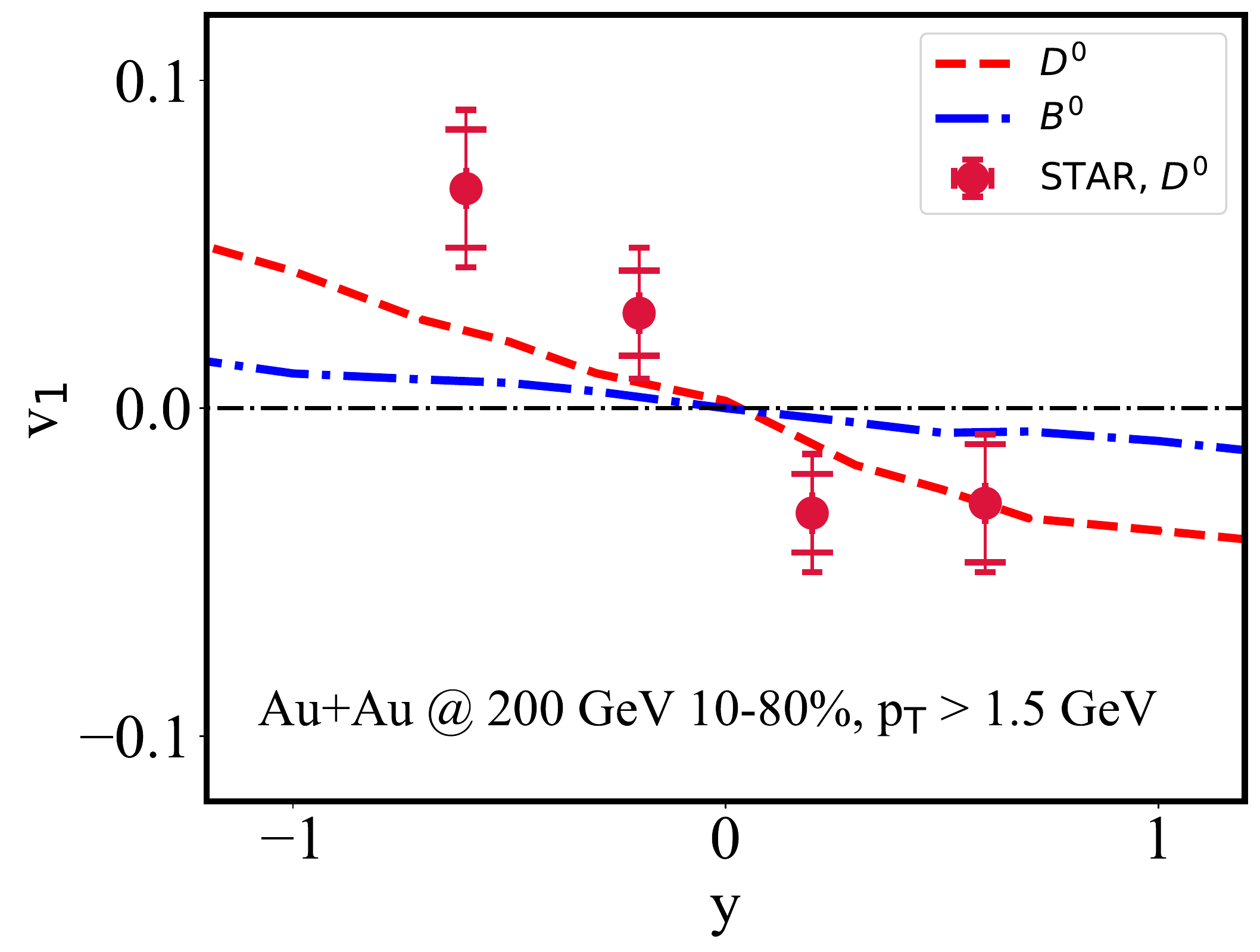}~\\
\end{center}
\caption{(Color online) Nuclear modification factor (upper panel), elliptic flow (middle panel) and directed flow (lower panel) of $D$ and $B$ mesons in 200~AGeV Au+Au collisions, compared to the STAR data~\cite{STAR:2014wif,STAR:2017kkh,STAR:2018zdy,STAR:2019clv}.}
\label{F1:RAA_AuAu}
\end{figure}

We start with the nuclear modification factor,
elliptic and directed flow coefficients of heavy mesons in 200~AGeV Au+Au collisions at RHIC.

In the upper panel of Fig.~\ref{F1:RAA_AuAu}, we present the $\raa$ of $D$ and $B$ mesons in 0-10\% Au+Au collisions in the mid-rapidity region ($|y|<1$). Using a spatial diffusion coefficient $D_\mathrm{s}(2\pi T)=4$ for $c$-quarks, our calculation provides a reasonable description of the $D$ meson $\raa$ measured by STAR~\cite{STAR:2014wif,STAR:2018zdy}. For $B$ mesons, a slightly smaller diffusion coefficient $D_\mathrm{s}(2\pi T)=3$ is used here, which is extracted from the $R_\mathrm{AA}$ of $b$-decay electrons ~\cite{STAR:2021uzu}. As discussed in the previous section, $D_\mathrm{s}(2\pi T)$ is treated as a model parameter here, whose detailed dependence on the heavy quark mass will be explored in a separate study soon. The peak structures of the $D$ and $B$ meson $\raa$ arise from the coalescence process that combines low $p_\mathrm{T}$ heavy and thermal light partons into medium $p_\mathrm{T}$ hadrons~\cite{Cao:2015hia}. Above the peak region ($\pt \gtrsim 2.5$~GeV), $B$ mesons exhibit a larger $\raa$ than $D$ mesons due to weaker energy loss of heavier partons through the QGP.

In the middle panel of Fig.~\ref{F1:RAA_AuAu}, we present the elliptic flow $v_{2}$ of $D$ and $B$ mesons in 10-40\% Au+Au collisions as a function of $\pt$ at mid-rapidity ($|y|<1$). Within the Langevin model using a constant $D_\mathrm{s}(2\pi T)$ value, our calculation underestimates the $D$ meson $v_2$ at its peak value ($\pt$ between 2 and 4~GeV) measured by STAR~\cite{STAR:2017kkh,STAR:2018zdy}. This indicates non-trivial dependences of the diffusion coefficient on the heavy quark momentum and the medium temperature, and can be improved with a more delicate calculation of the non-perturbative interactions between heavy quarks and the QGP at low $\pt$~\cite{Xing:2021xwc}. Here we also present the $v_2$ of $B$ mesons, which is non-zero but much smaller than that of the $D$ mesons. This is consistent with the findings observed in their $\raa$, suggesting weaker energy loss of $b$-quarks than $c$-quarks due to the larger mass of the former.

With the same diffusion coefficients used above, we present the rapidity dependence of the $D$ meson $v_1$ and predict the $B$ meson $v_1$ in 10-80\% Au+Au collisions in the lower panel of Fig.~\ref{F1:RAA_AuAu}. Our calculation qualitatively describes the trend of the $D$ meson $v_{1}$ observed at STAR~\cite{STAR:2019clv}. We see that both $D$ and $B$ mesons exhibit negative slopes of $v_1$ with respect to rapidity, due to the longer (shorter) path length of heavy quarks along the $+x$-direction than the $-x$-direction in the positive (negative) rapidity region, as illustrated in Fig.~\ref{fig:tiltedMedium}. This is a direct feature from a tilted QGP fireball. Since $b$-quarks are heavier than $c$-quarks, the unbalanced energy loss of $b$-quarks is smaller than that of $c$ quarks between $+$ and $-x$ directions, resulting in a smaller slope of the $B$ meson $v_1$ than the $D$ meson $v_1$. The slope parameters we extract around the $y=\pm 1$ regions are $dv_{1}/dy=-0.045\pm0.005$ for $D$ mesons and  $dv_{1}/dy=-0.010\pm0.002$ for $B$ mesons. 


\subsection{$\raa$, $v_{1}$ and $v_{2}$ of heavy flavor decayed electrons}
\label{section4-2}

The Heavy Flavor Tracker (HFT) at RHIC-STAR is able to measure single electrons from charm and beauty semi-leptonic decays~\cite{ATLAS:2020yxw}, providing a complementary tool to investigate properties of heavy quarks with particular species, considering the challenges in reconstructing $D$ and $B$ mesons in experiments. In this subsection, we present model calculations on the $R_\mathrm{AA}$, $v_1$ and $v_2$ of charm and beauty decayed electrons, and study how they depend on the medium geometry.

\begin{figure}[tbp]
\begin{center}
\includegraphics[trim=0cm 0.2cm 0cm 0cm,width=8 cm,height=5 cm,clip]{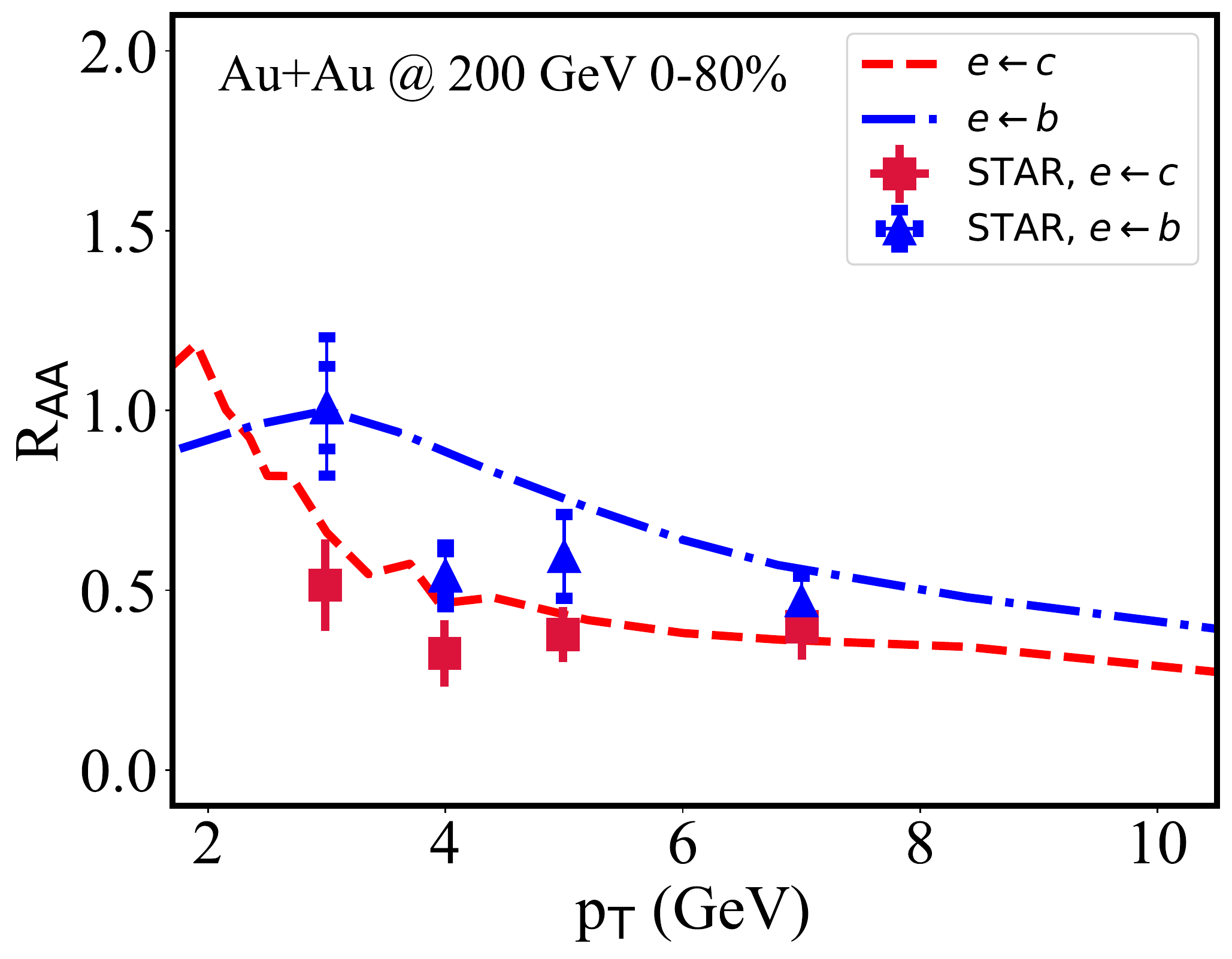}~\\
\includegraphics[trim=0cm 0.2cm 0cm 0cm,width=8 cm,height=5 cm,clip]{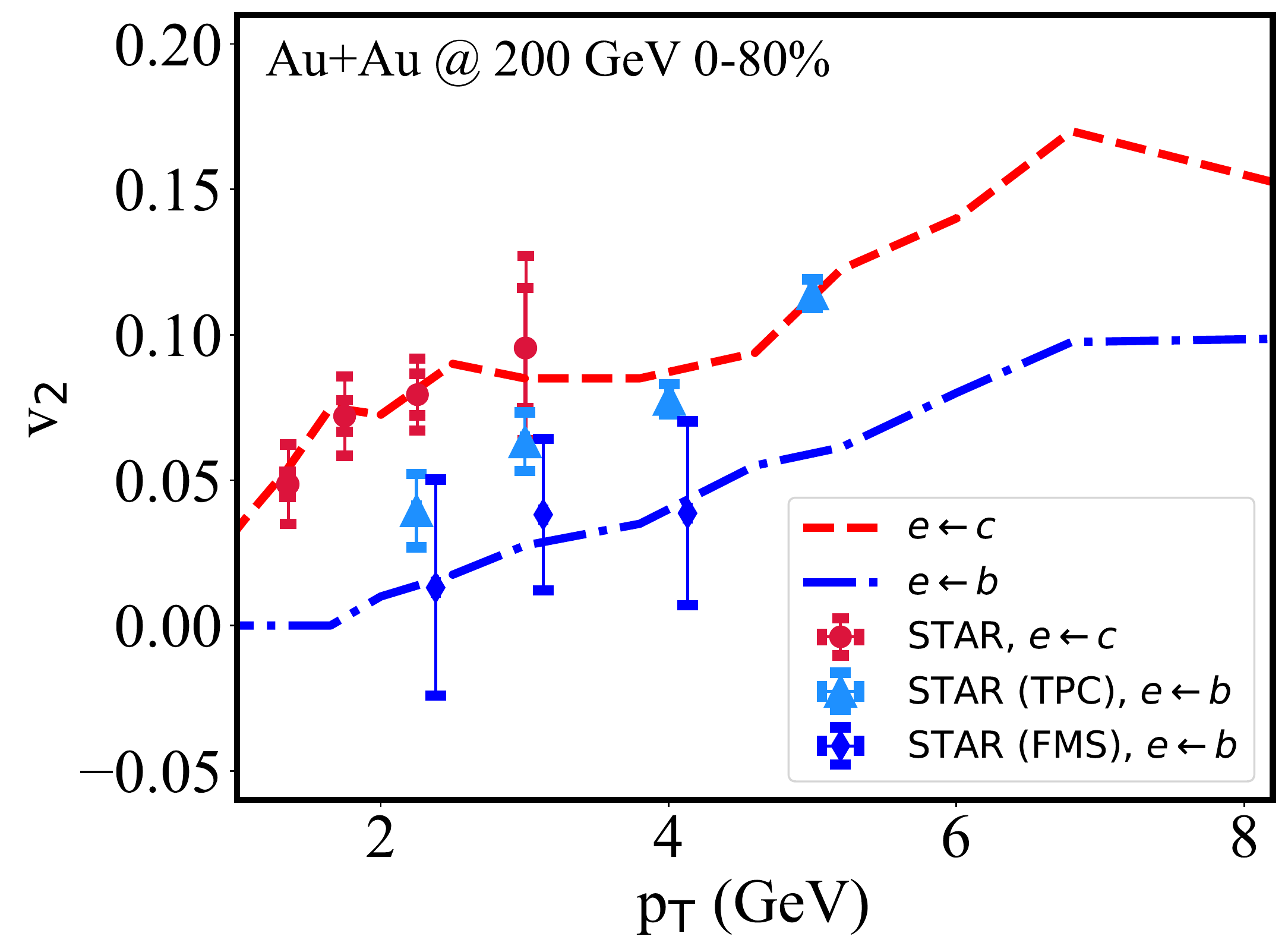}~\\
\includegraphics[trim=0cm 0.2cm 0cm 0cm,width=8 cm,height=5 cm,clip]{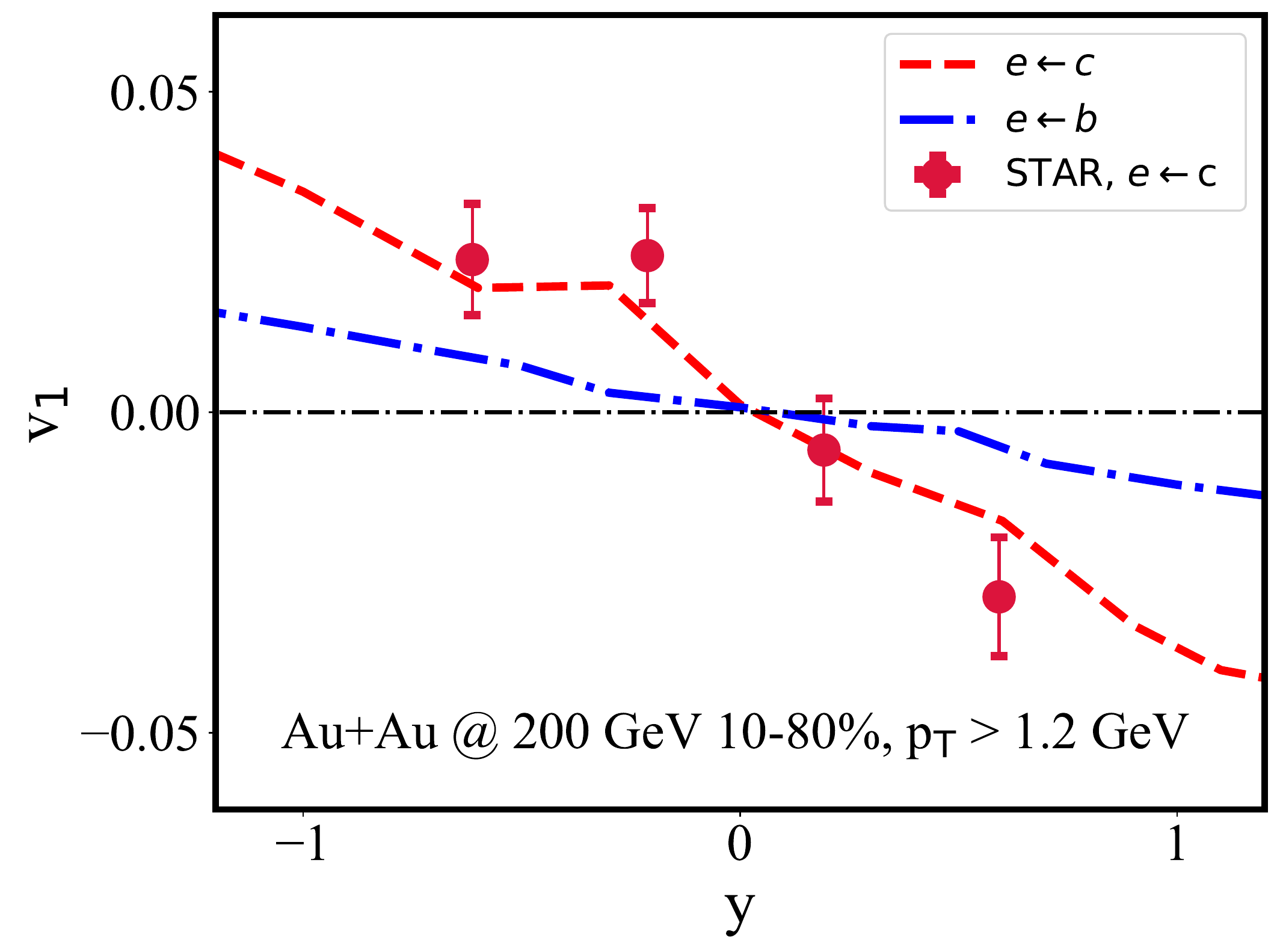}~
\end{center}
\caption{(Color online) Nuclear modification factor (upper panel), elliptic flow (middle panel) and directed flow (lower panel) of $c$ and $b$-decayed electrons in 200~AGeV Au+Au collisions, compared to the STAR data~\cite{Licenik:2020cjc,Kelsey:2020bms,Kramarik:2021emg,STAR:2021uzu}.}
\label{F:raa_electron}
\end{figure}

We start with the nuclear modification factors of $c$ and $b$-decayed electrons at mid-rapidity in the upper panel of Fig.~\ref{F:raa_electron}. With the same diffusion coefficients as we used in the previous subsection for heavy mesons ($D^{0}$ and $B^{0}$), our model calculation provides a reasonable description of $c$ and $b$-decayed electron $\raa$ observed by STAR~\cite{STAR:2021uzu} in 0-80\% Au+Au collisions at $\snn=200$~GeV. Note that although the flavor (or mass) hierarchy of parton energy loss is not obvious at high $\pt$~\cite{Xing:2019xae}, a clear difference between the charm and beauty quark energy loss can be observed here in the kinematic region focused by the RHIC experiment.

After fixing the diffusion coefficients with the heavy flavor $\raa$, we calculate their $v_2$ in the middle panel of Fig.~\ref{F:raa_electron}. Our model results of both $v_{2}(e\leftarrow c)(\pt)$ and $v_{2}(e\leftarrow b)(\pt)$ are in good agreement with the STAR measurement~\cite{STAR:2021uzu} for 0-80\% Au+Au collisions. A larger $v_2$ of charm decayed electrons than beauty decayed electrons is seen in both our model calculation and the experimental data, which is consistent with the hierarchy in their $\raa$ (the upper panel), and also that in the heavy meson $\raa$ and $v_2$ (Fig.~\ref{F1:RAA_AuAu}). Note that the discrepancy seen in the $D$ meson $v_2$ between our model result and the STAR data (middle panel of Fig.~\ref{F1:RAA_AuAu}) is not shown here for the electron $v_2$, indicating certain features of the heavy flavor dynamics may be shadowed by the momentum shift during the decay process.

The directed flow coefficient $v_{1}$ of $c$ and $b$-decayed electrons are shown in the lower panel of Fig.~\ref{F:raa_electron} as a function of rapidity. Our calculation provides a good description of the $c$-decayed electron $v_1$ measured by STAR~\cite{Kelsey:2020bms,Kramarik:2021emg} in 10-80\% Au+Au collisions at $\snn=$ 200 GeV, with a slope parameter extracted as $dv_{1} /dy = - 0.043\pm0.005$ around the $y=\pm 1$ regions. The $v_{1}(y)$ of $b$-decayed electrons is also predicted, with its slope parameter extracted as $dv_{1}/dy=-0.013\pm0.003$ around $y=\pm1$, which can be tested by future measurement at RHIC. The rapidity dependence of the heavy flavor decayed electrons here further confirms the longitudinally tilted geometry of the QGP fireball produced at the RHIC energy.

\subsection{Dependence of the heavy flavor $v_{1}$ on $\pt$}

While it is now generally accepted that the tilted geometry of the QGP generates the observed rapidity dependence of $v_1$ of $D$ mesons and their decayed electrons at RHIC, its $p_\mathrm{T}$ dependence has not been sufficiently discussed yet. This is the focus of this subsection.

\begin{figure}[tbp]
\begin{center}
\includegraphics[trim=0cm 0.2cm 0cm 0cm,width=8 cm,height=5 cm,clip]{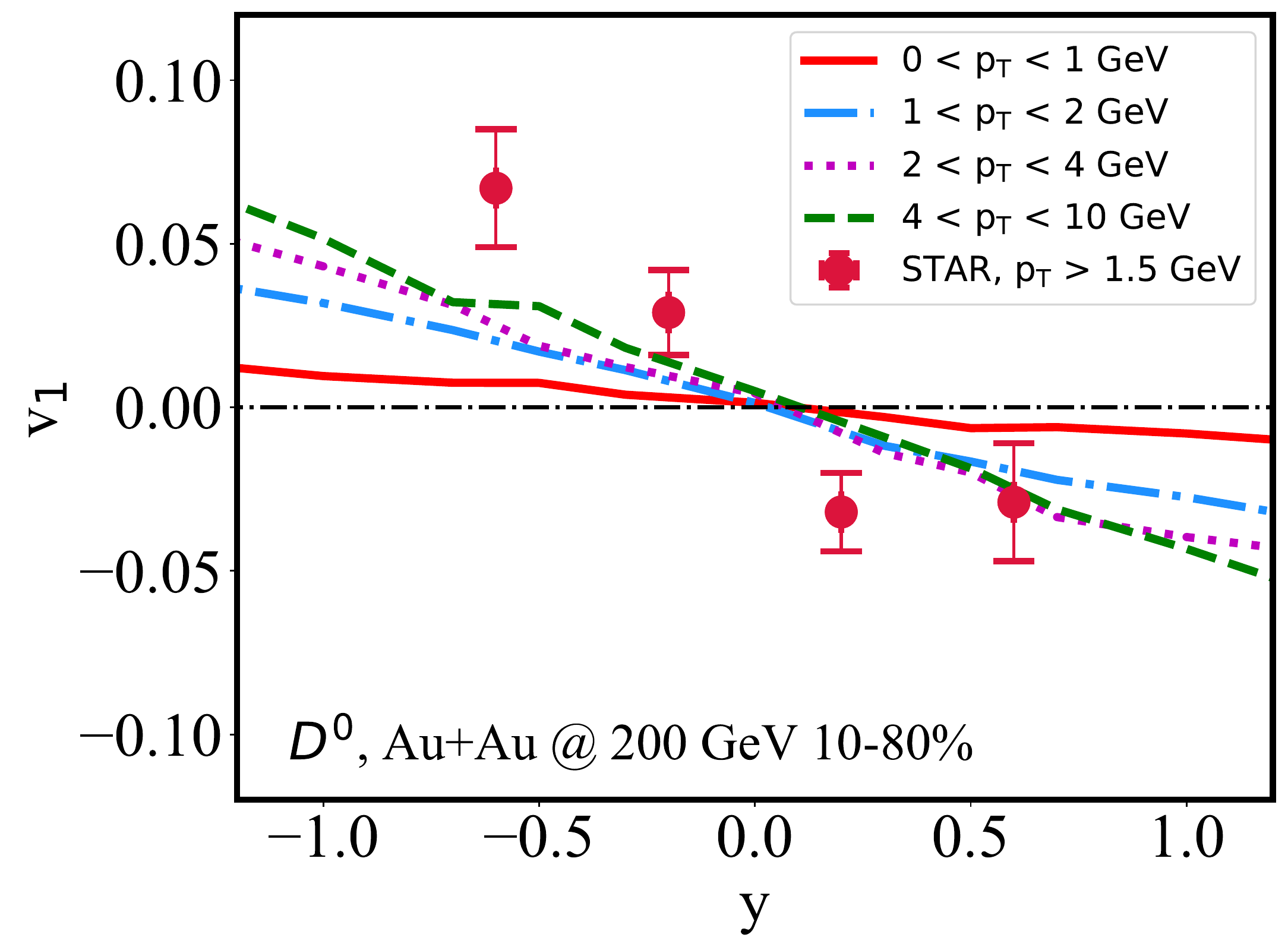}~\\
\includegraphics[trim=0cm 0.2cm 0cm 0cm,width=8 cm,height=5 cm,clip]{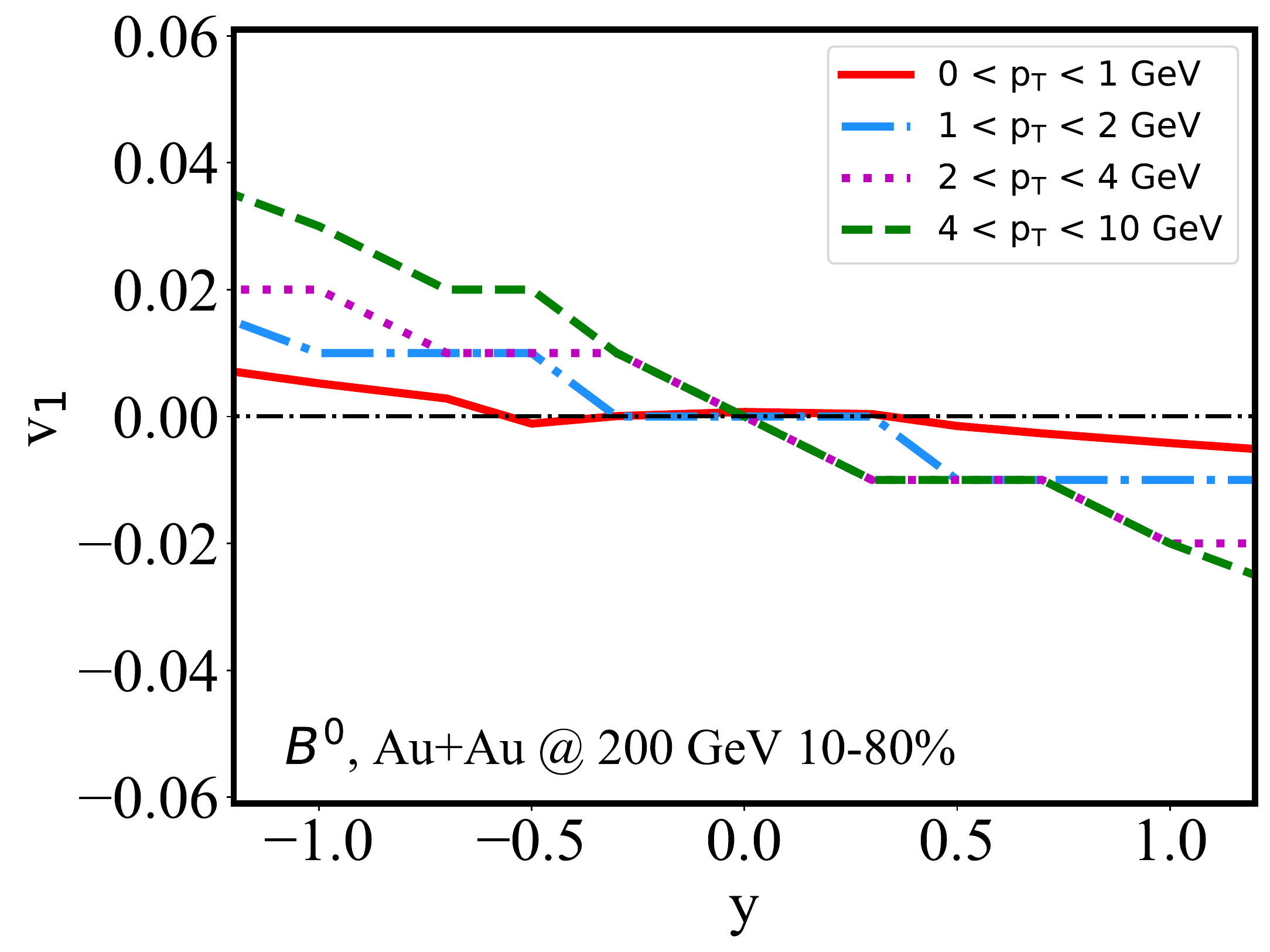}~
\end{center}
\caption{(Color online) (Color online) Directed flow coefficients of $D$ (upper panel) and $B$ (lower panel) mesons in different $\pt$ bins in 10-80\% Au+Au collisions at $\snn=200$~GeV.}
\label{pt_v1}
\end{figure}

\begin{figure}[tbp]
\begin{center}
\includegraphics[trim=0cm 0.2cm 0cm 0cm,width=8 cm,height=5 cm,clip]{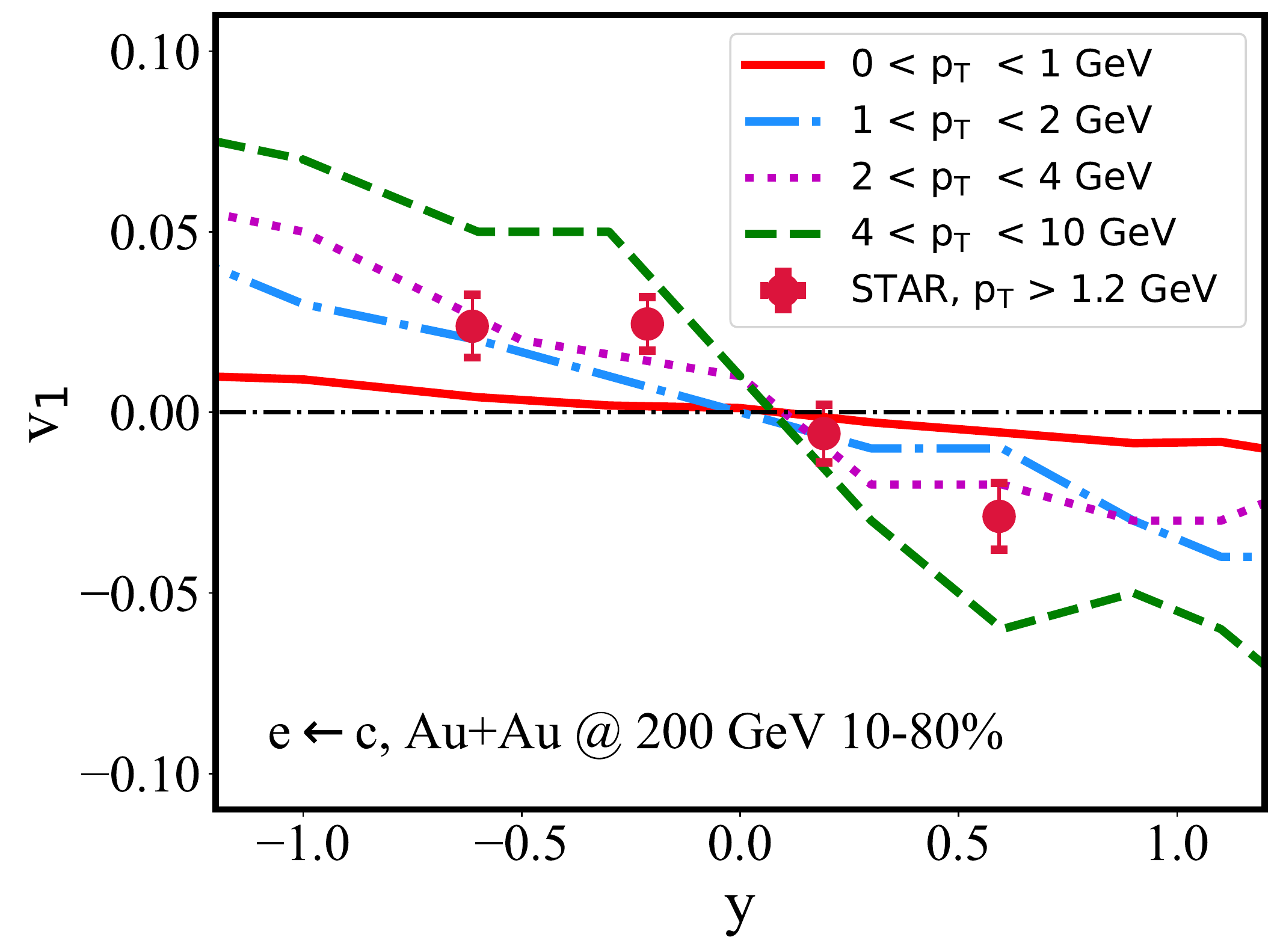}~\\
\includegraphics[trim=0cm 0.2cm 0cm 0cm,width=8 cm,height=5 cm,clip]{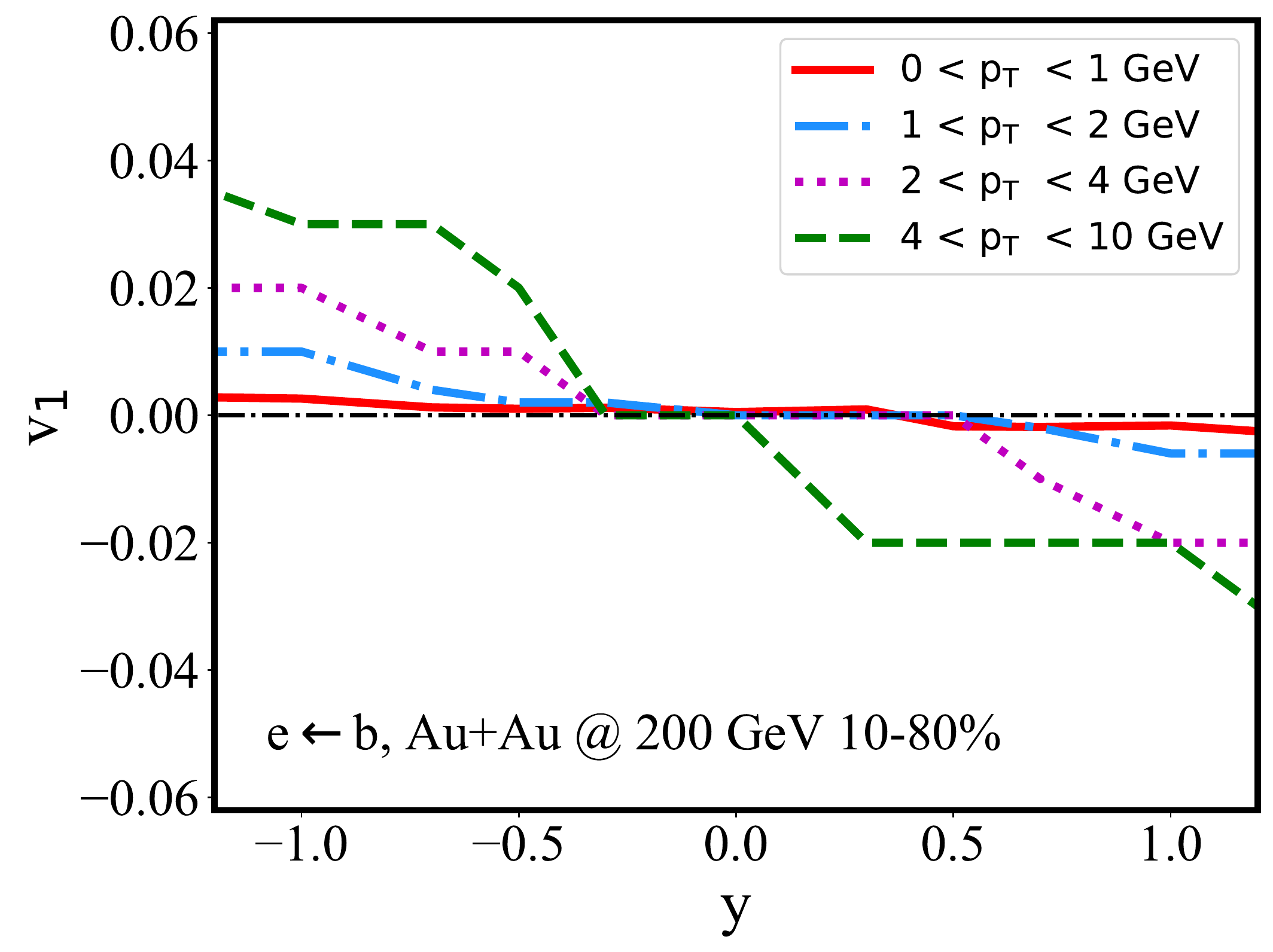}~
\end{center}
\caption{(Color online) Directed flow coefficients of charm (upper panel) and beauty (lower panel) decayed electrons in different $\pt$ bins in 10-80\% Au+Au collisions at $\snn=200$~GeV.}
\label{pt_v1_e}
\end{figure}

Shown in Fig.~\ref{pt_v1} is the $v_{1}$ of $D$ mesons (upper panel) and $B$ mesons (lower panel) for different $\pt$ regions in 10-80\% Au+Au collisions at $\snn = 200$~GeV. It is interesting to see that with the increase of $\pt$, the heavy meson $v_1$ becomes larger. This can be understood with the different origins of the heavy flavor $v_1$ at different $\pt$ scales. At very low $\pt$, heavy quarks tend to thermalize with the medium thus encode the thermal properties of the QGP. Since the $v_1$ of soft hadrons emitted from the QGP is small~\cite{Jiang:2021ajc}, one can expect a small $v_1$ of heavy quarks as well. 
Within the kinematic regions we explore here, the maximum slope parameter we obtain is for the $4<\pt<10 \textrm{~GeV}$ bin, whose value is extracted as $dv_{1}/dy=-0.050\pm 0.005~(-0.025\pm0.005)$ for $D$ ($B$) mesons around $y=\pm1$, which can be tested by future measurements.

The similar study is also conducted for the heavy flavor decayed electrons in Fig.~\ref{pt_v1_e}, upper panel for 
charm and lower panel for beauty decayed electrons in different $\pt$ bins of 10-80\% Au+Au collisions.
Consistent with the previous results for $D$ and $B$ mesons, we observe the slope of $v_{1}(y)$ becomes larger for higher $\pt$ bins. For $4<\pt<10$~GeV, the slope parameter is extracted as  $dv_{1}/dy=-0.065\pm0.005~(-0.025\pm0.005)$ for $c$ ($b$)-decayed electrons around $y=\pm1$.

\subsection{Nuclear modification factor along different directions}

\begin{figure}[tbp]
\begin{center}
\includegraphics[trim=0cm 0.2cm 0cm 0cm,width=8 cm,height=5 cm,clip]{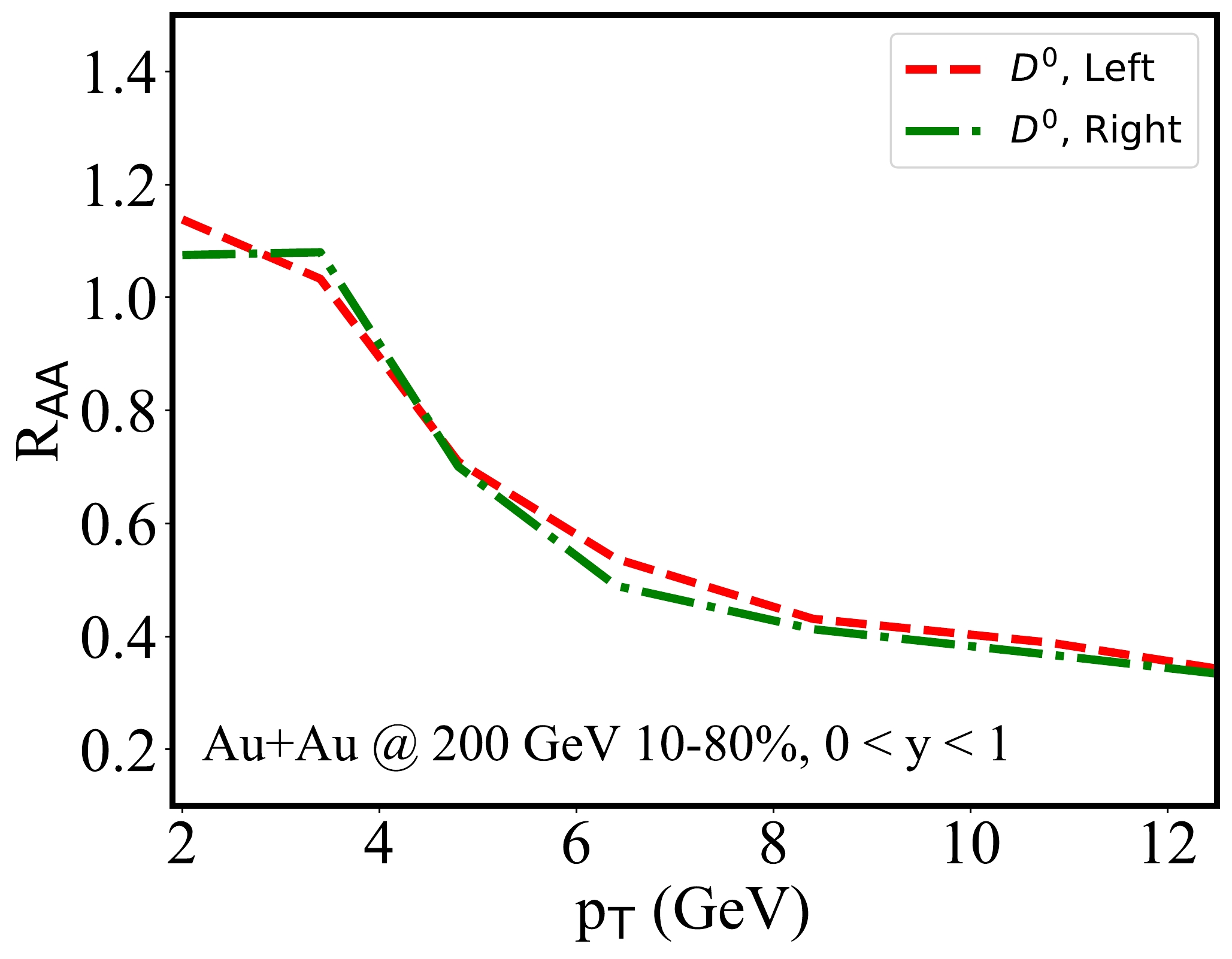}~ \\
\includegraphics[trim=0cm 0.2cm 0cm 0cm,width=8 cm,height=5 cm,clip]{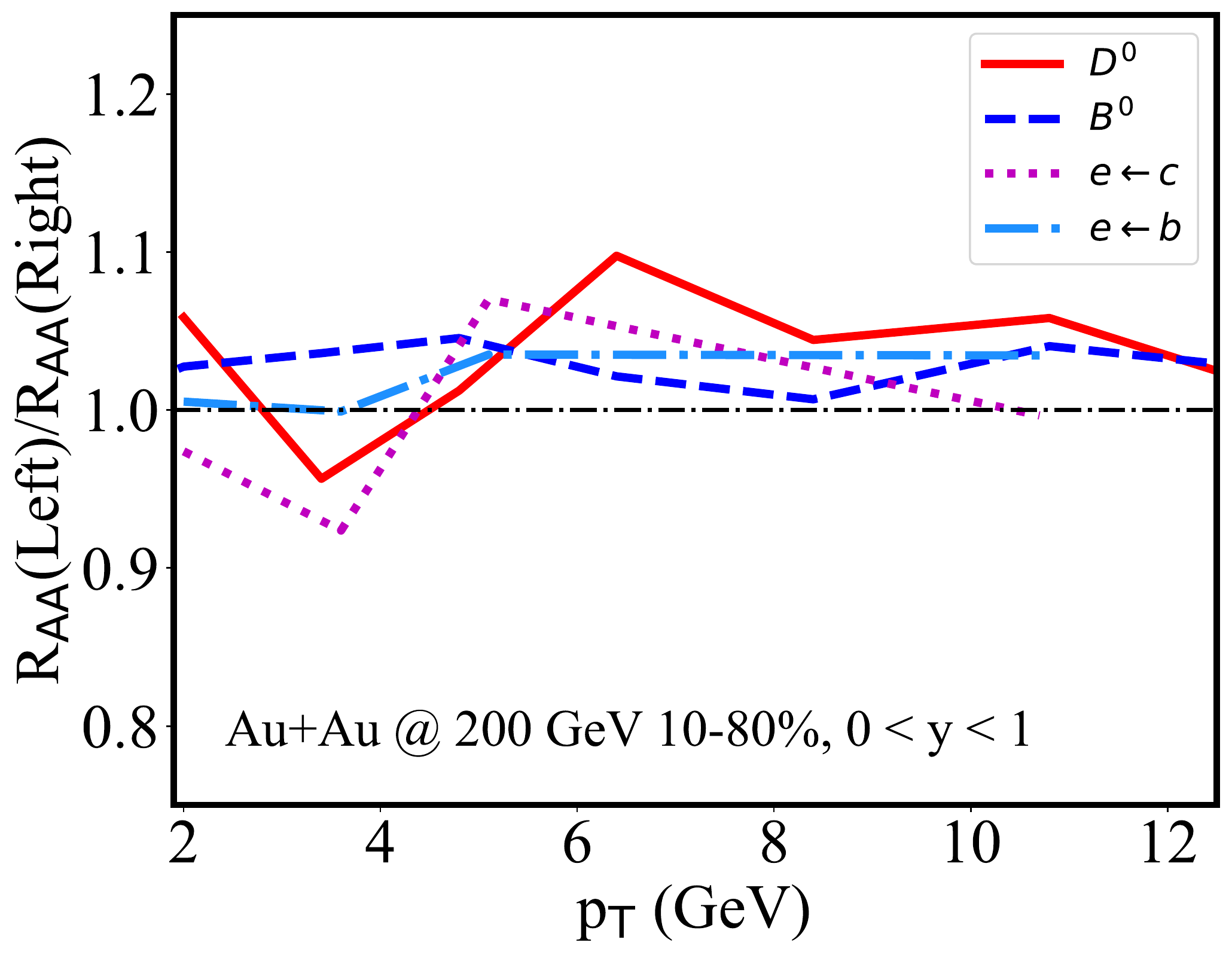}~\\
\end{center}
\caption{(Color online) The heavy flavor $\raa$ in $+$/$-x$ (right/left) hemispheres (upper panel) and the ratios between them (lower panel) in 10-80\% Au+Au collisions at $\snn=200$~GeV.}
\label{F:raa_asy}
\end{figure}

In addition to the directed flow coefficient, an alternative way to quantify the asymmetry of energy loss along different directions is studying the angular dependence of the $\raa$~\cite{Adil:2005qn,Jia:2009tf,Jia:2010ee}. In this subsection, we close our study by comparing the heavy flavor $\raa$ in different angular regions.

As previously illustrated in Fig.~\ref{fig:tiltedMedium}, in the positive rapidity region of a tilted QGP medium, heavy quarks that are initially produced symmetrically around the origin (0,0) propagate through a longer and hotter medium towards the $+x$ (right) direction than towards $-x$ (left), thus lose more energy in the right hemisphere than in the left.
Therefore, the azimuthal angle dependence of the heavy flavor $\raa $ can be utilized to investigate the violation of the longitudinal boost invariance in nuclear collisions.  We measure the azimuthal angle counter-clockwise from the $+x$ direction, and call the $-\pi/2<\phi<\pi/2$ region as the ``right" region and $\pi/2<\phi<3\pi/2$ as ``left".
In the upper panel of Fig.~\ref{F:raa_asy}, we present the $D$ meson $\raa$ as a function of its $\pt$ in the forward rapidity region of 10-80\% Au+Au collisions at $\snn=200$~GeV, analyzed within the left and right regions separately. Indeed, at high $\pt$ ($\gtrsim 4$~GeV), we observe a smaller $\raa$(right) than $\raa$(left) of $D$ mesons in the $0<y<1$ region. This could be viewed as an alternative signal of the longitudinal tilted fireball produced at the RHIC energy. At lower $\pt$ ($\lesssim4$~GeV), the heavy meson $\raa$ is also strongly affected by the coalescence process that is sensitive to the radial flow of the QGP, thus may not directly reflect the energy loss asymmetry of heavy quarks towards different directions.

To better illustrate the heavy flavor $\raa$ in different angular regions, we present the ratio between $\raa$(left) and $\raa$(right) in the lower panel of Fig.~\ref{F:raa_asy}. Results are shown for $D$ and $B$ mesons, as well as their decayed electrons. One observes that above $\pt\sim 4$~GeV, these $\raa$ ratios are all consistently above one in the forward rapidity region.

\section{Summary and outlook}
\label{section5}

In this paper, a systematic investigation on the heavy flavor nuclear modification factor ($\raa$), directed flow ($v_{1}$) and elliptic flow ($v_{2}$) is presented.
Effects from a longitudinally tilted bulk medium on these observables have been explored within a modified Langevin transport model coupled to a (3+1)-D viscous hydrodynamic model CLVisc.

Within this framework, our calculation provides a reasonable description of the $\raa$, $v_1$ and $v_2$ of heavy mesons and their decayed electrons compared to data currently available at RHIC. A clear mass hierarchy of parton energy loss can be observed within the $\pt$ range focused by RHIC experiments, where $D$ mesons exhibit smaller $\raa$ but larger $v_1$ and $v_2$ than those of $B$ mesons. The same hierarchy remains between charm and beauty decayed electrons. We have demonstrated that $v_{1}$ and the angular-dependent $\raa$ of heavy mesons and their decayed electrons encode information of the initial longitudinal deformation of the QGP energy distribution. An initially counter-clockwise tilted QGP fireball in the $\eta_\mathrm{s}$-$x$ plane results in a positive (negative) $v_1$ of heavy quarks in the backward (forward) rapidity regions. In addition, at high $\pt$, a smaller heavy flavor $\raa$ in the $+x$ (right) hemisphere than in $-x$ (left) is proposed at forward rapidity, which serves as an alternative observable to help constrain the 3D geometry of the QGP profile. The opposite conclusion is expected at backward rapidity. Furthermore, dependence of the directed flow on the heavy flavor $\pt$ has been studied in this work at the RHIC energy. As $\pt$ increases, heavy quarks become less thermalized and their observables are more dominated by their energy loss through the QGP. Within the kinematic range we have explored in this work, the heavy flavor $v_1$ increases as a higher $\pt$ region is applied. Our conclusions here consistently hold across $D$ and $B$ mesons and their decayed electrons, and await test from future experimental observations. 


While this work contributes to a more comprehensive understanding of how the heavy flavor probes can be utilized to constrain the initial geometry of the nuclear matter, it can be further extended in several directions. For instance, the recent isobar experiments at RHIC provide a novel environment to study the properties of QGP produced by colliding nuclei with the same number of nucleons but different geometries~\cite{STAR:2021mii,Xu:2017zcn,Xu:2021vpn,Jia:2022qrq}. It would be interesting to investigate whether this difference in nuclear structure can also be probed using heavy flavor observables within our framework. In addition, although the effect of the electromagnetic field on heavy quarks is weak at the RHIC energy, it becomes a dominating factor generating the heavy flavor $v_1$ at the LHC energy~\cite{Jiang:2022uoe}. However, it still remains a challenge to establish an ideal spacetime evolution profile of the electromagnetic field for a precise description of the heavy meson $v_1$ observed at LHC~\cite{Das:2016cwd,Chatterjee:2018lsx,Oliva:2020doe,Sun:2020wkg}. These will be explored in our follow-up efforts.

\begin{acknowledgements}
We are grateful for helpful discussions with Jiaxing Zhao, Xiang-Yu Wu and Guang-You Qin. This work was supported by the National Natural Science Foundation of China (NSFC) under Grant Nos.~11935007, 12175122 and 2021-867, Guangdong Major Project of Basic and Applied Basic Research No.~2020B0301030008, the Natural Science Foundation of Hubei Province No.~2021CFB272, the Education Department of Hubei Province of China with Young Talents Project No.~Q20212703, the Open Foundation of Key Laboratory of Quark and Lepton Physics (MOE) No.~QLPL202104 and the Xiaogan Natural Science Foundation under Grant No.~XGKJ2021010016.
\end{acknowledgements}

\bibliographystyle{unsrt}
\bibliography{heavy_v1ref}

\end{document}